\documentclass[12pt]{article}
\input{epsf.tex}
\begin{document}
\begin{center}
{\bf PARAMETERS OF THE $^{125}$Te COMPOUND STATE CASCADE $\gamma$-DECAY}\\
\end{center}
\begin{center}
{ V.A. Bondarenko$^1$,
J. Honz\'atko$^2$, V.A. Khitrov$^3$, A.M. Sukhovoj$^3$, I. Tomandl$^2$}\\
\end{center}
\begin{center}
{\it $^1$Institute of Solid State Physics, University of Latvia, LV 2169\\
 $^2$Nuclear Physics Institute, CZ-25068, \v{R}e\v{z} near Prague, Czech Republic\\
$^3$FLNP, Joint Institute for Nuclear Research, Dubna, Russia\\}
\end{center}
\begin{abstract}
Reliable information on level density $\rho$ and radiative strength functions
$k$ for the excitation energy region
with density of excited states  $\approx 10^2 MeV^{-1}$ and higher can be 
obtained now only by its model-free extraction from intensities $I_{\gamma\gamma}$
of two-step cascades proceeding between compound states and few low-lying 
levels. 
Full model-free determination of $\rho$ and $k$ in any method is possible only if one
can extract from additional experimental information about general trend in dependence of
ratio of strength functions of emitted reaction products of a given type on
excitation energy $E_{ex}$ of the nucleus under study.

Analysis of the available experimental data  for $^{125}Te$ shows that the  peculiarities
observed earlier in other  nuclei are also inherent to this nucleus.
\end{abstract}

\section{Introduction}

Level density $\rho$ and strength functions of emission the nuclear reaction products
are, first all, the test 
for any nuclear models. Naturally, confidence level of  experimental values of these parameters should be 
high enough. If confidence of experimental level density in the regions of neutron resonanses and low-lying 
levels is extremely high than the contrary situation is observed out of this region of nuclear excitation.
The same can be said about the radiative strength functions $k$ of the primary gamma-transitions of the 
compound states decay and strength functions $S$ of emission of products in different nuclear reactions. 
This discrepancy is completely caused by different ways of determination $\rho$ in the excitation energy 
regions where level spacing $D$ is significantly larger or less than the resolution $\sigma$ of spectrometers 
used in experiment. In the case  $\sigma>> D$ and  $\rho$, and $k$ should be determined only 
simultaneously with the use of mathematics apparatus for solution of reverse problems. In the case under 
consideration, however, they don not have simple solution even in principle.

The situation gets complicated by that, for the long time, the only method to determine level density was its 
extraction from the spectra of evaporated nucleons in different nuclear reactions. Besides, this method 
requires the use of theoretical strength functions for emission of nucleons or light nuclei. They are to be 
determined in the frameworks of nuclear models with a precision exceeding required accuracy in 
determination of  $\rho$. 

Up to now, comparison between the experimental and calculated strength functions for reactions like $(d,p)$ 
or $(d,t)$ was performed up to the excitation energy of about a half of the neutron binding energy $B_n$ 
(see, for example, [1]). It shows that the details of fragmentation process of any states of nuclear potential 
over real levels can be reproduced in calculation within modern nuclear models only with the error of about 
several hundreds percents. This is caused by both approaches of conventional models and uncertainties of 
their parameterization. The scale of above errors follows, first of all, from un-removable and very significant 
discrepancy between the experimental and calculated energies even for the most low-lying levels of the 
simplest structure.

Similar problems arise and in the analysis of the gamma-ray spectra from any nuclear reaction where 
product nucleus is excited up to 5-8 MeV and higher. In spite of existence of the advanced enough models of 
the radiative strength functions, the accuracy of the predicted by them $k$
values primary transitions with low energy 
$E_\gamma$ is really unknown. This requires one to develop new, model independent methods for 
simultaneous determination $\rho$ and $k$ values with high enough precision.

In this case the most serious problem is the necessity to extract information about general trend in  
dependence  of strength functions ratio on  excitation energy $E_{ex}$ for given type of emitted reaction 
products from complementary experiment. This problem is  partially solved [2] only in analysis of the 
experimental  intensities $I_{\gamma\gamma}$ of two-step cascades proceeding between compound states 
and low-lying  levels and gamma-ray intensities in single spectra measured with $Ge$ detectors. Earlier 
version [3] of this  method used zero assumption about independence of energy dependence of strength 
functions on the energy $E_{ex}$ of decaying level. Of course, this led to additional (although small enough) 
systematic errors of determined $\rho$ and $k$.

That is why, one can conclude:   reliable information on level density $\rho$ and radiative strength functions 
$k$ for the excitation energy region with density of excited states  $\approx 10^2 MeV^{-1}$ and higher 
can be  obtained now only in the frameworks of method [2]. 
The advantages of method [2] over other known methods to determine level density [4] and radiative 
strength functions [5] are stipulated by the following:

a) the dependence of $I_{\gamma\gamma}$ on absolute level densities (in other methods, the spectra 
do not depend on absolute values of  $\rho$ and $k$);

b) the smallest systematic errors of   two-step cascade spectra
$\delta I_{\gamma\gamma}$ as compared with
spectra of all known methods of determination of $\rho$ and $k$;

c) much smaller transfer coefficients of systematic errors of measured
    intensities to the parameter errors $\delta \rho$ and $\delta k$;

d) practically unique fixation of the spin interval of excited levels; 

e) although functional dependence of the two-step cascade intensity on level density and radiative 
strength functions cannot be simple but the interval of possible variations of $\rho$ and $k$ is 
always limited by several tens percent under condition that the $I_{\gamma\gamma}$ value used in 
[2] is [6] is the function of only of energy of the cascade primary transition $E_1$.

\section{Possibility to estimate the  shape of dependence $k(E_\gamma, E_{ex})$}

Comparison of intensities $i_{\gamma\gamma}$ of the two-step
cascades proceeding through given intermediate levels $E_i$ with independently determined [7]
intensities of the primary $i_1$ and secondary $i_2$ transitions allows determination of cascade
population $P-i_1= (i_1 \times  i_2/ i_{\gamma\gamma})-i_1$ for a set of levels of any nucleus up to the 
excitation energy $2-4$ MeV and higher.  In practice, its error cannot exceed the total error in determination 
of $i_2$  ($\approx 20-25\%$ even for relatively weak cascades or gamma-transitions).
This parameter is very sensitive to the shape of dependence of $k(E_\gamma , E_{ex})$
on $E_{ex}$. Moreover, the ratio $k(E_\gamma ,E_{ex})/k(E_\gamma , B_n)$ for gamma-transitions with 
equal energy and multipolarity,  to the first approach, can be simply enough estimated [2] from calculation 
of population within different  level density and strength function models.

There is no possibility to determine population of all without exclusion cascade intermediate levels $E_i$ 
even at low excitation owing to the presence of registration threshold for intensities $i_{\gamma\gamma}$,  $i_1$ and  $i_2$.
The Porter-Thomas fluctuations of the primary transition intensities and possible dependence of population of levels on 
structure of their wave functions complicate comparison between the experimental and calculated  values of 
$P$ for individual cascades and choice of adequate  to the experiment models and hypotheses.

It is worthwhile to compare cascade populations summed in small intervals of excitation energy. These sums 
should be considered as the lower estimations for every interval.

The extent of discrepancy between the calculated population $P-i_1$ and its lover estimation  is determined 
by both incompleteness of data on intensities of cascades and transitions and possible strong influence of 
structure of wave function of excited level on probability of its cascade population. This permits on to 
estimate [2] the dependence of $k$ on $ E_{ex}$.

\section{Systematic errors in determination of experimental population of levels}

To achieve high confidence level of $\rho$ and $k$ derived [2] from cascade intensities with the use of 
cascade population of levels it is necessary:

 to get large enough set of experimental data on the intensities $i_{\gamma\gamma}$,  $i_1$ and  $i_2$    with small enough 
errors;

 to determine the quantum ordering in cascades with minimum number of false assignments.

Extraction of individual cascades as pairs of resolved peaks with minimum registration threshold is affected 
by the following sources of systematic errors:

a) random grouping of events in resulting spectrum (result of background extraction) in pairs of         
false peaks;

b) registration of three-step cascades with pair of detectors in full-energy peaks.

The minimum statistics error in determination of peak areas in spectra with fixed total energy 
$E_1+E_2=B_n-E_f$ is provided by the use numerical method [8] for improving resolution without 
decreasing of efficiency. As a result, each spectrum is symmetrical with respect of its center. In the 
frameworks of obvious assumption that the probability of formation of random peaks in the region of 
negative or positive number of events is equal than analysis of part of spectrum with negative number of 
events provides rather objective determination of minimal cascade intensities which guarantee very small 
probability to obtain false value of $i_{\gamma\gamma}$.

A part of background events relates to registration of cascade with higher total energy so that one quantum is 
registered in the full-energy peak and another -- in Compton background. Application of method [8] leads to 
formation of specific sign-variable structures with zero mathematics expectation of their areas.  They may 
mask true cascades.  Identification and minimization [9] of total area of these structures does not find any 
difficulties because determination of parameters of intense cascade with higher energy is simple, as well.

Probability of registration of any cascade quantum with $E_\gamma \geq 0.5$ MeV in full-energy peak in performed
experiment does not exceed 1\% per 1 emitted gamma-quantum.  Three-step cascades total intensity
(in sum with two-step cascades with one pure quadrupole 
quantum) can be estimated from part of events corresponding to possible cascades to final levels with spin 
difference  $|J_i-J_f|=3$. In all 50 nuclei studied by us it does not exceed several tenths of percent per decay.

Due to multitude of variants of gamma-transition energies in three-step cascades, these events form wide 
spectrum of peaks of some amplitude. This amplitude is inversely proportional to number of variants. That is 
why, maximum absolute intensity of three-step cascade with given energy of the third quantum, most 
probably, cannot exceed $\sim 10^{-4}$ of decays. The effect quickly decreases as increasing the energy 
$E_f$ of level populated by the pair of previous transitions. Therefore, distortions of the obtained $P-i_1$ values 
owing to registration of three-step cascades and corresponding increase in $i_{\gamma\gamma}$ are negligible in the nuclear 
excitation energy region $ E_{ex} > 2-3$ MeV. But just this region demonstrates maximum influence of 
nuclear structure on the shape of energy dependence of radiative strength functions. 

Some difficulties in determination of $P-i_1$ are made by unresolved in traditional [7] experiment doublets of 
cascade secondary transitions $i_2$. Difficulties related to doublets of primary transitions are some less due to 
less density of peaks in high-energy part of corresponding spectra. Partially, multiplets can be identified and 
resolved within the approximation procedure of single $HPGe$ detector spectra when studying the thermal 
neutron radiative capture with the use of information on two-step cascades and evaluated decay schemes. 
Besides, intensity $i_2$ can be distributed between cascades -- members of multiplets --
proportionally to $i_{\gamma\gamma}$.
But 
it is preferable to exclude such events from the procedure of determining 
$P$ if the data on $i_{\gamma\gamma}$,  $i_1$ and  $i_2$ for 
any cascade intermediate level are superfluous.

Another problem is made by overlapping of peaks in  $\gamma$-ray spectra following thermal neutron radiative 
capture, especially in the region in the vicinity of $E_\gamma \approx 0.5B_n$. This leads to additional underestimation of 
intensities $i_1$, $i_2$ and population of level. This underestimation increases when energy of the level under 
consideration increases, too.

Spectroscopic data derived from cascade intensities have the highest confidence because the statistics stored 
provides  true variant of the Ritz combinatorial principle  at the level of 98-99\% or more. In practice, all events of 
false coincidences of gamma-quanta sum energies of different cascades with the difference of level energies 
are excluded from data acquisition by electronics.

The main uncertainty in the obtained decay schemes in nuclei with any level density is related with 
impossibility of experimental determination of life time of cascade intermediate level and, respectively, 
quantum ordering in it. The maximum likelihood method [10] for its determination cannot provide errorless 
choice of position of cascade intermediate level from two variants. Minimization of this error can be 
achieved only at accounting for all the complex [11] of spectroscopic information stored for the nucleus 
under study.

All the data used for determination of $P$ values are listed in table.
%In Table 1 are given only 
%cascades with intermediate levels whose existence does not call doubts [7,11] with the high probability. The 
%second contains cascades with earlier unknown levels whose energy $E_i =B_n-E_1$  is determined under 
%condition $E_1 >E_2$. But inverse quantum ordering for them is most probable.

\section{Reproduction of experimental level population in calculations} 

Experimental and calculated level populations can be compared in two variants:

the total population of each of  $N_i$ intermediate levels (including intensities of populating them 
primary transitions) is compared (Fig.~1) with several variants of calculation;

experimental results are compared (Fig.~2) with  cascade population summed over 200 keV 
excitation energy intervals of cascade intermediate levels.

The necessity of the use of both variants is stipulated, first of all, by the registration threshold for cascades 
which limits experimental information on possible levels of a nucleus under study. Besides, the errors in 
determination of $i_{\gamma\gamma}$,  $i_1$ and  $i_2$ sometimes results in negative values of $P-i_1$
(this strongly manifests itself at 
small as compared with i1 cascade population of level). Moreover, the total population $P$ of levels depends 
on model values of level densities and radiative strength functions sufficiently less than the cascade 
population $P-i_1$. This occurs owing to compensation of effect on population of decrease, for instance, in 
$\rho$ by increase in $k$ for the cascade primary transitions. 
(Total gamma-width of compound state is constant value).

Therefore, comparison between the experiment and model calculation only for summed $P-i_1$ values cannot 
give complete picture of the process. Complementary but qualitative confirmation for significant 
discrepancy between the calculation and experiment is provided by comparison of total population of levels.

There is a number of energy dependencies of strength functions and level densities being suitable for 
calculation of $P-i_1$. But general regularities of change in level population as changing level excitation energy 
can be revealed using only three variants of calculation:

(a)	the level density is described by any (for example, [12]) model of non-interacting Fermi-gas, 
strength function for $E1$ transitions is set by known [13,14] extrapolations of the giant electric 
dipole resonance into the region below $B_n$ and $k(M1)$=const with normalization of the ratio 
$k(M1)/k(E1)$ to the experiment in the vicinity of $B_n$;

(b)	the calculation uses $\rho$ and  $k$ values obtained according to [3] and providing precise description of 
energy dependence of two-step cascade intensities (at present – only for the cascade final levels with 
$E_f <1$ MeV);

(c)	 in the calculation there are involved level densities and strength functions providing simultaneous 
and precise ($\chi^2/f<<1$) description of $I_{\gamma\gamma}=F(E_1)$ (Fig.~3), total radiative width 
$\Gamma_\gamma$of decaying 
compound state and total cascade populations $P-i_1$.

Variant (c) can be realized in iterative regime:  for the data obtained according to [3], there is chosen some 
functional dependence which changes strength functions for secondary transitions with respect to strength 
function [3] so that better to reproduce $P-i_1$ values. For this, it is quite enough to multiply strength functions 
of secondary gamma-transitions to the levels below some fixed energy  by the function $h$ which contains 
several narrow peaks. Dependence of their form on the nuclear excitation energy can be determined by 
analogy with specific heat of ideal microsystems in the vicinity of point of the second order phase transition 
as the following:

\begin{equation}
h=1+\alpha\times ({ln(|U_c-U_1|)-ln(|U_c-U|)})~~if~~U<U_c,
\end{equation}
\begin{equation}
h=1+\alpha\times ({ln(|U_c-U_2|)-ln(|U_c-U|)})~~if~~U>U_c,
\end{equation}
with some parameters $\alpha,~U_1,~U_2$ $U_c$.

 Condition  $(U_c-U_1) \neq (U_2-U_c)$ provides asymmetry of peaks and more 
precise description of cascade population as compared, for instance, with the Lorentz curve. 
In the best variant tested by us, the amplitude $\alpha$ must grow  from zero
(linearly, for example) up to the maximally
possible value shown on fig. 4 as the excitation energy $U$ decreases from $U=B_n$ to
$U=U_c$. 
Positions of peaks, their form and amplitude are determined by the $P-i_1$ values. Unfortunately, more 
precise notions about the shape of energy dependence of radiative strength functions of given multipolarity  
cannot be derived from this analysis. 

Population of the level $l$ is calculated according to equation
\begin{equation}
P_l=\sum_m P_m\times \Gamma_{m,l}/\Gamma_m,
\end{equation}
It depends on population $P_m$ of all the higher-lying levels $m$ and branching ratio at their decay. Although 
the data in figs.1 and 2 depend on both multiplicands in eq.(3) but $P_l$ values for different low-lying levels are 
mainly determined by ratio of partial widths $\Gamma_{m,l}$ of populating them secondary transitions. Equation (3) 
cannot provide other possibility for significant increase of calculated population  of high-lying levels when 
population of low-lying level decreases.

The obtained corrective functions are then involved in analysis [3] to derive $\rho$ and $k$ parameters which allow 
precise reproduction of cascade intensities with accounting for assumed difference in energy dependence of 
strength functions for primary and secondary cascade transitions. Values of $i_{\gamma\gamma}$  are shown in Fig.~3, re-
determined level densities and radiative strength functions are presented in figs. 4 and 5. If it is necessary, 
this procedure is repeated one times with the use of the hypothesis of linearly increasing distortions of the 
$k(E1)$ and $k(M1)$ values when energy of decaying levels decreases and several times for hypothesis $h$=const. 
For minimization of number of fitted parameters, corrective functions in figs. 4 and 5 were supposed equal 
for both electric and magnetic gamma-transitions.

It should be noted, that the population of levels lying below ``step-like" structures in level density cannot be 
reproduced without assumption about decrease in corresponding radiative strength functions also in sum 
energy interval. On the whole, function $h$ at least qualitatively repeats the most general shape of 
dependence obtained according to [3], in practice, for all nuclei: significant increase in $k$ for gamma-
transitions to levels from the region of ``step-like" structures and decrease for gamma-transitions to lower-
lying levels. If these regularities do not have alternative (unknown) explanation, then simple extrapolation of 
the obtained results into the region $E_{ex} > B_n$  allows conclusion about possible analogous energy dependence 
of radiative strength functions and for primary gamma-transitions following fast neutron radiative capture. 
And, as a consequence, about necessity of both experimental test of models of radiative strength functions 
and level densities for these excitation regions and modification (to more or less extension) of algorithms for 
calculation and evaluation of corresponding cross-sections.

The open question is evident discrepancy between the experimental and calculated results in figs. 1 and 2. It 
cannot be removed in the frameworks of assumption about independence of the averaged  partial width of 
gamma-transitions populating some levels on structure of wave functions of these levels. This results from 
very big difference in calculated and experimental total populations $P$, for example, in the interval 3.0-3.5  
MeV.

For 30 observed here (Fig. 1) from 100 expected (Fig. 4) levels, experimental population is approximately 9 
times larger than the results of the best variant of calculation. This contradiction cannot be related only to the 
Porter-Thomas fluctuations of the widths of primary transitions populating the levels: probability of such or 
bigger random divergence $P$ for one level equals  $\sim 0.001$. The only realistic explanation is 
assumption about strong dependence of population of level on its structure. For example, nucleus may have 
two or more systems of levels of different structure. Then the mean values of $\rho$ and $k$ (figs. 4 and 5) 
cannot be considered as the mean arithmetic values. Conventional theoretical notions do not take into 
account this possibility. But it naturally follows from interpretation of the data presented in figs. 4 and 5.
	
A number the used hypotheses is inevitable at the achieved stage of the problem under solution. In this case, 
all the conclusions about the cascade gamma-decay process of compound state is to be considered, most 
probably, as qualitative than as quantitative. So, clearly expressed ``step-like" structure in level density and 
related with it increase in $k(E1)+k(M1)$ (Fig. 4) should be considered as established with high confidence 
level. But the number and shape of these ``steps", most probably, can be determined only in the further 
experiments. The same can be said about corrective functions $h$. If energy region corresponding to 
significant increase in $k$ for the second, third and so on cascade transitions (due to the number of tested 
variants) calls no doubts then parameters of function $h$ have to be considered, most probably,  as very 
preliminary. They should be used, first of all, for planning of corrective experiments.

In this sense, the region $h<1$ is very demonstrative. Whether the strength of gamma-transitions is re-
distributed from the lower-lying levels populated by them to higher-lying or this structure of $h$ provides 
only narrowing of width of the region of maximum increase in $k(E1)+k(M1)$ values in Fig. 4 -- nobody can 
answer this question. But it should be noted that we could not reproduce observed values $P-i_1$ for all 
available set of  nuclei without significant decrease  in $k$.

The fact that only the lower estimation of $P-i_1$ was obtained in the experiment cannot be possible 
explanation. So, at low excitation energy the difference the numbers of intermediate cascade levels and 
levels from evaluated decay schemes decreases (due to increase in the mean intensity of cascades at 
practically constant threshold of their registration). This means that experimental $P-i_1$  (Fig. 2) must increase 
with respect to values shown in fig. 2 as increasing excitation energy and decreasing in registration 
thresholds for $i_{\gamma\gamma}$,  $i_1$ and  $i_2$.
As a consequence, the difference in energy dependence of strength functions for 
the primary and secondary transitions must increase – strength functions of gamma-transitions to high-lying 
levels must intensify. 

Specific dependence of the product $k\times h$ (local peaks in the second multiplicand) qualitatively corresponds to 
theoretical [15] regularities of fragmentation of the state of any structure over nuclear levels. One of the 
important conclusion of this theoretical analysis is that the strength of an state ($N$ quasi-particles and $M$ 
phonons) concentrates in asymmetric peaks of limited width with "tails" in the region of high excitation 
energy.

\section{The most probable values of level density and radiative strength functions of cascade $\gamma$-transitions in   
$^{125}Te$}

The method described above provides determination of model-free, rather precise and confident 
values of $\rho$ and $k$. Unfortunately, in addition to some sources of possible systematic errors discussed above 
their values can include errors that are specific for different nuclei. For example, absolute value of $k$ can 
be distorted by local deviation of the neutron resonance density   from its general tendency owing to 
possible but not taking into account in calculated values of $I_{\gamma\gamma}$ structure effects.
Or owing to possible correlations of partial 
radiative widths of cascade transitions and reduced neutron width which determines the main part of neutron 
capture cross-section. Similar correlations can change the ratio between intensities of cascades proceeding 
through intermediate levels with different structure and lead to additional error in $\rho$ and $k$.

Analysis [2] of the available experimental data [7,16] for $^{125}Te$ shows that the  peculiarities
observed earlier in other studied nuclei manifest themselves in this nucleus, as well.
There are:

two ``step-like"  structures in the energy dependence of $\rho$ and

correlating with their positions variations in shape of energy dependence of $k$.

The data on significant increase in radiative strength functions of secondary transitions, in practice, in the 
same energy interval as for primary transitions should be considered as one more additional confirmation for 
existence in a nucleus of the excitation energy region where occurs abrupt change of its structure. One can 
assume that there is transition from domination of vibrational type excitations to domination of quasi-
particle excitations. Apparently, this fact can be interpreted as the phase transition from super-fluid to 
normal state of such very specific system as a nucleus. Probably, this effect is related to breaking of the only 
nucleon pair [17] at the excitation energy corresponding to radical increase in level density.

\section{Conclusion}

The data on investigating properties of $^{125}$Te are in good agreement with analogous results obtained earlier. 
One can consider them as preliminary indication at possible phase transition related, most probably, with 
breaking of one Cooper pair in the region beside $0.5B_n$ and the next one approximately on 2 MeV higher. 
Quantitative information on intensifying strength functions of the secondary gamma-transitions in given 
excitation energy interval of the nucleus can be useful for planning more detailed experiments on direct 
investigations of dynamics of breaking of Cooper pairs in different nuclei. Unlike any known super-fluid 
macrosystems, nuclei are the limited and heterogeneous according to type of statistics and magnitude of 
inner energy of Cooper pair with respect to the Fermi energy.
\newpage
\begin{flushleft}
{\large\bf References}
\end{flushleft}
\begin{flushleft}

1. Honz\'atko J. et al., Nucl. Phys., 1999 {\bf 645} 331.\\
2. Bondarenko V.A. et all, In: XII International Seminar on \\\hspace*{14pt}Interaction
       of Neutrons with Nuclei, E3-2004-169, Dubna, 2004, p. 38.\\
%\hspace*{14pt} http://arXiv.org/abs/nucl-ex/0406030\\
3. Vasilieva E.V., Sukhovoj A.M., Khitrov V.A., Phys. At. Nucl., 2001 
{\bf 64(2)} 153.\\
 \hspace*{14pt} http://arXiv.org/abs/nucl-ex/0110017\\
4. Zhuravlev B.V. Bull.Rus.Acad.Sci.Phys., 1999 {\bf 63} 123.\\
5. Voinov A. et al. Phys. Rev., 2001 {\bf C 63} 044313.\\
6. Boneva S.T., Khitrov V.A., Sukhovoj A.M.,
 Nucl.  Phys. 1995, {\bf A589} 293.\\
7. Honz\'atko J. et al., Fizika B (Zagreb), 1998 {\bf 7} 87.\\
8. Sukhovoj A.M.,  Khitrov V.A., Instrum. Exp. Tech.,  1984 
{\bf 27}
1071.\\
9.	Bondarenko V.A., Honz\'atko J., Khitrov V.A., Sukhovoj  A.M., Tomandl I.,
 Fizika B \hspace*{14pt} (Zagreb), 2002 {\bf 11} 201.\\
10. Popov Yu.P.,  Sukhovoj A.M.,  Khitrov V.A.,  Yazvitsky Yu.S.
,
Izv. AN SSSR,\\\hspace*{14pt}  Ser. Fiz., 1984 {\bf 48} 1830.\\
11. http://www.nndc.bnl.gov/nndc/ensdf.\\
12. Dilg  W., Schantl W., Vonach  H., Uhl  M., Nucl. Phys. 1973, V. A217, P. 269-298.\\
13. Kadmenskij S.G., Markushev V.P. ,  Furman W.I., Sov. J. Nucl. Phys., 1983 {\bf 
37}  165.\\
14. Axel P., Phys. Rev., 1962 {\bf 126(2)} 671-683.\\\hspace*{14pt} 					           Blatt J. M.,  
Weisskopf V. F. Theoretical Nuclear Physics. New York (1952).\\
15. Malov L.A., Soloviev V.G., Yad. Fiz.,  1977 {\bf 26(4)}  729.\\
16. Honz\'atko J., Tomandl I., Khitrov V.A., Sukhovoj A.M.,Ninth International\\\hspace*{14pt} 
Symposium on Capture  Gamma-Ray  Spectroscopy and Related Topics, \\ 
\hspace*{14pt}Budapest1996, Ed. G.L.Molnar, T.Belgya, Zc. Revay, Springer, 1997,  p.438.\\
17. Ignatyuk A.V., Sokolov Yu.V., Yad. Fiz., (1974) {\bf 19} 1229.\\
\end{flushleft}

\newpage
{\sl Table 1.\\ A list of absolute intensities $I_1$, $I_s$ for cascade quanta,
$i_{\gamma\gamma}$ is two-step cascade intensities with energies $E_1$ and $E_2$, of
the cascade transitions, $E_i$ is the energy of the intermediate levels.
$E_s$  - possible secondary cascade quanta energy.}
\begin{center}
\begin{tabular}{|l|r|r|r|r|r|c|}  \hline
$I_1$,\%&  $E_i$, keV &$E_2$, keV &
$i_{\gamma\gamma}$ & $E_s$, keV & $I_s$, \% &M\\\hline
   1.15&   443.5(4)&   443.5&   590(17)&  443.53&   8.80&   \\
       &             &   408.0&   584(26)&  408.04&   5.20&   \\
       &   463.4(4)&   463.4&      11(3)&  463.35&   2.30&   \\
       &             &   427.9&      35(7)&  427.85&   6.40&   \\
  0.31&   537.8(4)&   537.8&   196(12)&  537.79&   5.80&   \\
       &             &   502.3&   107(29)&  502.3&   3.30&  2\\
       &   671.2(3)&   635.7&      14(4)&  635.91&   4.20&   \\
  1.80&   729.0(4)&   729.0&   260(11)&  729.22&   1.59&   \\
       &             &   693.5&   480(18)&  693.72&   5.73&   \\
       &  1016.9(4)&   981.4&      5(2)&  981.7&   0.08&   \\
       &  1054.5(7)&  1054.5&      7( 2)&       &        & \\
       &             &  1019.0&      8( 2)&       &        & \\
  0.10&  1054.5(7)&   611.0&      46(13)&  610.22&   0.94&   \\
  0.08&  1064.6(15)&  1064.6&      13( 3)&       &        & \\
       &             &  1029.1&      6(2)& 1029.3&   0.07&   \\
       &             &   539.2&      19( 6)&       &        & \\
       &  1071.6(4)&   546.2&      16(5)&  546.56&   3.10&   \\
  0.07&  1241.6(8)&  1241.6&      12(3)& 1241.1&   0.12&   \\
       &             &  1206.1&      16(4)& 1207.29&      1.12&   \\
  0.14&  1264.7(4)&  1264.7&      8(3)& 1264.91&   0.08&   \\
       &             &  1229.2&      77(7)& 1229.67&   0.87&  2\\
       &  1315.4(3)&   852.1&      19(5)&  851.9&   0.04&   \\
  3.70&  1319.0(3)&  1319.0&   169(11)& 1319.5&   0.36&  2\\
       &             &  1283.5&   162(10)& 1284.2&   0.45&  2\\
       &             &   855.7&   224(13)&  856.18&   0.36&  2\\
       &             &   793.6&   1788(51)&  794.22&   3.02&   \\
       &             &   381.1&        (1)&       &        & \\
  0.14&  1528.7(7)&  1493.2&      23(5)& 1493.3&   0.78&  3\\
       &             &  1085.2&      33(4)& 1086.1&   0.29&   \\
       &             &  1065.4&      32(3)& 1066.29&   0.66&  2\\   
       &             &  1119.5&      7( 3)&       &        & \\ 
  0.23&  1587.2(12)&  1587.2&   120(8)& 1587.27&   0.73&   \\
       &             &  1551.7&      16(3)& 1551.8&   0.08&   \\
       &             &  1143.7&      33(4)& 1143.9&   0.09&   \\
       &             &  1123.9&      14(3)& 1123.3&   0.06&   \\
       &             &  1061.8&      08( 3)&       &        & \\
       &             &   858.0&      15(2)&  858.5&   0.04&   \\ 
 0.09&  1652.5(9)&  1114.7&      13(3)& 1115.2&   0.07&   \\
       &             &   923.3&      23(3)&  923.29&   0.29&   \\
       &             &   331.4&      16( 6)&       &        & \\
\hline \end{tabular} \end{center}
 \newpage%\vspace*{9.8mm}
\begin{center}
\begin{tabular}{|l|r|r|r|r|r|c|}  \hline
$I_1$,\%&  $E_i$, keV &$E_2$, keV &
$i_{\gamma\gamma}$ & $E_s$, keV & $I_s$, \%& M\\\hline   
  0.07&  1669.6(2)&  1669.6&      28(4)& 1669.9&   0.31&   \\
       &             &  1131.8&      18(3)& 1132.4&   0.25&   \\
       &             &   998.2&      10(2)&  998.5&   0.11&   \\
       &             &   940.4&      19(3)&  940.9&   0.16&  2\\
  0.22&  1713.0(3)&  1713.0&      25(4)& 1713.4&   0.37&  2\\
       &             &  1677.5&      20(5)& 1678.2&   0.15&   \\
       &             &  1269.5&      58(8)& 1269.9&   0.23&   \\
       &             &  1175.2&      8(3)& 1175.5&   0.21&  3\\
       &             &  1041.6&      18(3)& 1042.1&   0.19&   \\
       &             &   983.8&      55(4)&  984.33&   0.29&   \\
       &  1767.8(3)&  1230.0&      11(3)& 1229.67&   0.87&  2\\
       &  1824.4(10)&  1788.9&      10(4)& 1788.0&   0.20&  2\\
       &             &  1286.6&      10(3)& 1286.6&   0.13&  2\\
       &             &   503.3&   243(64)&  502.3&   3.30&  2\\
       &  1829.8(13)&  1829.8&      15(5)& 1829.7&   0.22&   \\
       &             &  1794.3&      15(5)& 1795.1&   0.19&  2\\
       &             &  1386.3&      16(4)& 1385.8&   0.11&   \\
       &             &  1292.0&      9( 3)&       &        & \\
  0.14&  1863.6(12)&  1863.6&      66(9)& 1863.6&   0.38&  2\\
       &             &  1828.1&      14(4)& 1827.3&   0.10&   \\
       &             &  1420.1&      12(5)& 1418.89&   0.60&  2\\
       &             &  1421.4&      49(9)& 1421.6&   0.17&  2\\
  0.08&  1898.8(4)&  1863.3&      24(5)& 1863.6&   0.38&  2\\
       &             &  1455.3&      24(5)& 1455.4&   0.22&   \\
       &             &  1452.9&      18(5)& 1452.9&   0.11&   \\ 
  0.15&  1904.3(3)&  1904.3&      38(7)& 1905.4&   0.38&   \\
       &             &  1868.8&      10(4)& 1868.3&   0.15&   \\ 
       &             &  1460.8&      42(5)& 1461.36&   0.35&   \\
       &             &  1366.5&      25(8)& 1367.2&   0.16&   \\
       &             &  1175.1&      19(3)& 1175.5&   0.21&  3\\
       &  1913.6(4)&  1878.1&      14(4)& 1878.5&   0.11&   \\
       &  1918.7(9)&  1918.7&      23(6)& 1919.6&   0.22&  2\\
       &             &  1883.2&      12(4)& 1882.5&   0.13&   \\
  4.20&  1956.0(2)&  1956.0&   2226(49)& 1956.73&   2.80&   \\
       &             &  1920.5&   181(18)& 1921.6&   0.42&   \\
       &             &  1492.7&   613(60)& 1493.3&   0.78&  3\\
       &             &  1418.2&   512(32)& 1418.89&   0.60&  2\\
  0.62&  1978.0(2)&  1978.0&   245(17)& 1978.76&   0.50&  2\\
       &             &  1942.5&      28(7)& 1942.8&   0.10&   \\
       &             &  1440.2&   220(22)& 1440.94&   0.49&   \\
       &             &  1306.6&      15(4)& 1307.26&   0.55&  2\\
  1.80&  2008.7(3)&  2008.7&   1145(36)& 2009.3&   1.50&  2\\
       &             &  1973.2&   100(13)& 1973.9&   0.19&   \\
       &             &  1470.9&      58(12)& 1470.6&   0.03&   \\
       &             &  1337.3&      78(9)& 1338.1&   0.21&  2\\    
\hline \end{tabular} \end{center} \newpage%\vspace*{9.8mm}
\begin{center}
\begin{tabular}{|l|r|r|r|r|r|c|}  \hline
$I_1$,\%&  $E_i$, keV &$E_2$, keV &
$i_{\gamma\gamma}$ & $E_s$, keV & $I_s$, \%& M\\\hline   
       &             &   936.8&   281(16)&  937.47&   0.30&  2\\
       &  2011.1(3)&   939.2&      40(11)&  940.9&   0.16&  2\\
       &  2043.6(5)&  2008.1&      14(5)& 2009.3&   1.50&  2\\
  0.04&  2049.0(4)&  1605.5&      22(7)& 1606.0&   0.27&   \\
       &             &  1319.8&      16(4)& 1319.5&   0.36&  2\\
  0.22&  2060.4(3)&  2024.9&      13(5)& 2025.5&   0.08&   \\
       &             &  1616.9&   152(14)& 1617.51&   0.50&   \\
       &             &  1522.6&      24(6)& 1522.4&   0.08&   \\
       &             &  1331.2&      12( 4)&       &        & \\
  0.25&  2076.3(3)&  2076.3&   129(13)& 2077.0&   0.49&  2\\
       &             &  2040.8&      20(5)& 2041.5&   0.16&   \\
       &             &  1538.5&      12(4)& 1538.7&   0.07&   \\
       &             &  1347.1&      15(4)& 1347.9&   0.07&   \\
       &             &  1004.4&      30(5)& 1005.2&   0.07&   \\
  0.67&  2107.8(5)&  2107.8&   108(12)& 2108.4&   0.25&   \\
       &             &  2072.3&      89(10)& 2073.1&   0.27&   \\
       &             &  1664.3&   107(13)& 1664.7&   0.12&   \\
       &             &  1570.0&      12( 4)&       &        & \\
       &             &  1035.9&   287(13)& 1036.73&   0.39&   \\
  0.50&  2129.0(2)&  2093.5&   217(15)& 2094.1&   0.56&  2\\
       &             &  1685.5&   162(16)& 1686.2&   0.36&   \\
       &             &  1591.2&      43(7)& 1591.8&   0.12&   \\
       &             &  1057.1&      22(5)& 1057.5&   0.09&   \\
       &  2172.0(5)&  2172.0&      12(4)& 2173.8&   0.06&   \\
 0.04&  2177.6(4)&   856.5&      27(7)&  856.18&   0.36&  2\\
 0.06&  2181.4(4)&  1737.9&      25(7)& 1738.4&   0.24&   \\
 0.05&  2203.0(6)&  1131.1&      24(8)& 1130.5&   0.06&   \\
 0.08&  2221.6(20)&  2186.1&      19(6)& 2186.7&   0.36&  2\\
       &             &  1758.3&      15(5)& 1756.6&   0.13&   \\
       &             &  1492.4&      30(9)& 1493.3&   0.78&  3\\
       &             &   900.5&      25( 7)&       &        & \\
  0.61&  2225.4(2)&  2189.9&      72(9)& 2190.7&   0.21&  3\\
       &             &  1782.0&      76(11)& 1782.8&   0.13&   \\
       &             &  1762.2&      50(8)& 1763.4&   0.09&   \\
       &             &  1554.1&   147(12)& 1554.8&   0.36&   \\
  0.07&  2246.0(4)&  2246.0&      13(4)& 2245.3&   0.11&   \\
       &             &  1174.1&      19(5)& 1175.5&   0.21&  3\\
       &  2259.6(5)&  2224.1&      13( 5)&       &        & \\
       &             &   938.5&      34(7)&  937.47&   0.30&  2\\
  0.09&  2270.6(5)&  2235.1&      14(5)& 2235.6&   0.26&   \\
       &  2282.8(5)&  2282.8&      13(4)& 2283.2&   0.13&  2\\
  0.23&  2314.3(11)&  2314.3&      24( 5)&       &        & \\
       &             &  2278.8&      16(5)& 2278.2&   0.07&   \\
       &             &  1776.5&      39(7)& 1778.0&   0.10&   \\
       &             &  1242.4&      34(5)& 1242.92&   0.44&   \\
       &  2332.1(3)&  2332.1&      17(4)& 2330.3&   0.17&  2\\
\hline \end{tabular} \end{center} \newpage%\vspace*{9.8mm}
\begin{center}
\begin{tabular}{|l|r|r|r|r|r|c|}  \hline
$I_1$,\%&  $E_i$, keV &$E_2$, keV &
$i_{\gamma\gamma}$ & $E_s$, keV & $I_s$, \%& M\\\hline   
       &  2338.8(4)&  2303.3&      17(5)& 2302.6&   0.13&   \\
       &             &  1017.7&      27(7)& 1018.36&   0.31&   \\
  0.25&  2351.5(7)&  2316.0&      37(7)& 2316.2&   0.11&   \\
       &             &  1680.1&      25(5)& 1680.1&   0.19&   \\
       &             &  1279.6&      18( 5)&       &        & \\
  0.34&  2379.1(2)&  2379.1&      37(6)& 2380.0&   0.25&   \\
       &             &  1841.3&      43(11)& 1841.5&   0.10&   \\
       &             &  1707.7&      19(5)& 1708.9&   0.12&   \\
       &             &  1649.9&      78(14)& 1650.3&   0.17&   \\
       &             &  1307.2&      14(5)& 1307.26&   0.55&  2\\
  0.05&  2403.3(6)&  2403.3&      11(4)& 2402.9&   0.32&  2\\
       &             &  2370.0&      49(7)& 2370.2&   0.23&  2\\
  0.06&  2409.5(2)&  1337.6&      33(5)& 1338.1&   0.21&  2\\
  0.07&  2414.4(4)&  2414.4&      12(4)& 2415.3&   0.08&   \\
       &             &  1743.0&      12(4)& 1744.1&   0.18&   \\
       &             &  2379.7&      20(6)& 2378.1&   0.07&  2\\
  0.05&  2438.6(4)&  2403.1&      20(6)& 2402.9&   0.32&  2\\
  0.26&  2466.2(2)&  2466.2&      68( 8)& 2467.03&   0.13&   \\
       &        (2)&  1928.3&   175(21)& 1928.9&   0.45&   \\
       &             &  2430.7&      77(11)& 2430.4&   0.13&   \\
       &             &  2022.7&   100(29)& 2022.9&   0.13&   \\
       &             &  1794.8&      92(9)& 1795.1&   0.19&  2\\
       &  2489.7(2)&  1417.8&      39(7)&       &        & \\
  0.04&  2503.7(4)&  2468.2&      16(5)& 2469.6&   0.11&   \\
       &  2520.8(11)&  2520.8&      18(5)& 2521.8&   0.23&   \\
       &             &  2485.3&      18(5)& 2485.8&   0.05&   \\
       &             &  2077.3&      31(8)& 2077.0&   0.49&  2\\
       &             &  1448.9&      22(6)& 1448.16&   0.08&   \\
       &  2529.4(4)&  2493.9&      15(5)& 2495.4&   0.19&   \\
       &  2543.4(4)&  2543.4&      21( 5)&       &        & \\
  0.46&  2549.5(4)&  2549.5&      84(9)& 2550.8&   0.39&  3\\
       &             &  2514.0&      63(10)& 2514.6&   0.12&   \\
       &             &  2106.0&      65(12)& 2106.3&   0.13&   \\
       &             &  2011.7&      72(12)&       &        & \\
       &             &  1477.6&      23(6)& 1477.8&   0.05&  2\\
       &             &  1228.4&      74(12)&       &        & \\
       &  2554.6(4)&  2554.6&      22(5)& 2554.7&   0.15&  2\\
  0.10&  2560.5(3)&  2560.5&      32(6)& 2560.7&   0.10&   \\
  0.41&  2566.7(13)&  2566.7&      68(8)& 2566.1&   0.10&   \\
       &             &  2531.2&   121(13)& 2532.8&   0.24&  2\\
       &             &  2123.2&      83(12)& 2122.4&   0.085&  2\\
       &             &  2103.4&      37(9)& 2103.4&   0.12&   \\
       &             &  2028.9&      22( 7)&       &        & \\
       &             &  1494.8&      25(6)& 1494.7&   0.37&  2\\
\hline \end{tabular} \end{center} \newpage%\vspace*{9.8mm}
\begin{center}
\begin{tabular}{|l|r|r|r|r|r|c|}  \hline
$I_1$,\%&  $E_i$, keV &$E_2$, keV &
$i_{\gamma\gamma}$ & $E_s$, keV & $I_s$, \%& M\\\hline 
  0.49&  2584.9(6)&  2584.9&      24( 5)&       &        & \\
       &             &  2549.4&   108(12)& 2550.8&   0.39&  3\\
       &             &  2141.4&      96(13)& 2141.8&   0.23&   \\  
       &             &  2121.6&      20(7)& 2122.4&   0.085&  2\\
       &             &  2047.1&      68(11)& 2047.3&   0.41&   \\
       &             &  1513.0&      29(6)& 1513.3&   0.47&   \\
  0.32&  2605.8(4)&  2162.3&      34(9)& 2163.2&   0.07&   \\
       &             &  2068.0&      18(6)& 2068.0&   0.15&   \\
       &             &  2606.3&   111(16)& 2607.0&   0.19&  2\\
       &             &  2570.8&      42(8)& 2571.9&   0.11&  2\\
       &             &  1534.4&      78(9)& 1535.0&   0.11&   \\
       &             &  1285.2&      49(13)& 1284.2&   0.45&  2\\
       &             &  2609.0&      60(16)& 2609.6&   0.45&  2\\
       &             &  1287.9&      34(12)& 1286.6&   0.13&  2\\
       &  2641.8(5)&  1320.7&      35(11)& 1322.8&   0.19&   \\
  0.87&  2649.0(7)&  2649.0&      18( 5)&       &        & \\
       &             &  2613.5&      55(9)& 2614.3&   0.17&   \\
       &             &  2205.5&      17(6)& 2204.8&   0.10&   \\
       &             &  2185.7&   299(17)& 2184.7&   0.08&   \\
       &             &  2111.2&      26(7)& 2111.4&   0.10&   \\
       &             &  1977.6&      24(9)& 1978.76&   0.50&  2\\
       &             &  1919.8&      57(13)& 1919.6&   0.22&  2\\
       &             &  1577.1&   177(14)& 1578.0&   0.28&   \\
  0.21&  2672.9(2)&  2672.9&      97(11)& 2674.2&   0.35&  2\\
  0.12&  2675.1(4)&  2639.6&      31( 7)&       &        & \\
       &             &  2137.3&      21( 7)&       &        & \\
       &             &  2213.9&      18(5)& 2215.9&   0.10&   \\
  0.13&  2690.2(10)&  2690.2&      15( 5)&       &        & \\
       &             &  2654.7&      16( 6)&       &        & \\
  0.12&  2705.1(2)&  2669.6&      58(9)& 2670.6&   0.19&  2\\
       &  2726.0(5)&  2282.5&      24(6)& 2281.2&   0.10&   \\
  0.16&  2728.6(10)&  2728.6&      54(8)& 2730.6&   0.31&  4\\
       &             &  2693.1&      16(6)& 2692.8&   0.19&   \\
       &             &  2285.1&      34( 7)&       &        & \\
       &             &  2190.8&      14(5)& 2190.7&   0.21&  3\\
  0.07&  2751.0(4)&  2751.0&      18(6)& 2751.1&   0.07&  2\\
       &             &  2715.5&      16(6)& 2716.3&   0.12&   \\
  0.08&  2754.0(4)&  2310.5&      19(6)& 2310.6&   0.08&   \\
       &             &  2216.2&      25(6)& 2218.4&   0.10&  2\\
  0.44&  2770.4(2)&  2770.4&      80(10)&       &        & \\
       &             &  2734.9&      60(8)& 2735.5&   0.15&   \\
       &             &  2232.6&      41(7)& 2233.0&  0.08&   \\
       &             &  1698.5&      81(11)& 1698.5&   0.30&   \\
  0.10&  2775.7(2)&  2332.2&      54(8)& 2333.1&   0.31&  2\\
\hline \end{tabular} \end{center}  \newpage%\vspace*{9.8mm}
\begin{center}
\begin{tabular}{|l|r|r|r|r|r|c|}  \hline
$I_1$,\%&  $E_i$, keV &$E_2$, keV &
$i_{\gamma\gamma}$ & $E_s$, keV & $I_s$, \%& M\\\hline 
  0.44&  2785.7(4)&  2785.7&      54(9)& 2784.3&   0.09&  2\\
       &             &  2750.2&      25(6)& 2751.1&   0.07&  2\\
       &             &  2247.9&      24(6)& 2247.2&   0.13&  2\\
       &             &  1713.8&      35(8)& 1713.4&   0.37&  2\\
       &             &  1464.6&      44(13)&       &        & \\  
  0.07&  2791.1(5)&  2755.6&      11( 4)&       &        & \\
     5&  2801.9(5)&  2766.4&      13( 5)& 2767.3&   0.05&  2\\
     7&  2813.8(10)&  2370.3&      21( 6)&       &        & \\
       &             &  2276.0&      18(6)& 2275.7&   0.08&   \\
  0.24&  2818.8(4)&  2783.3&      35(6)& 2784.3&   0.09&  2\\
       &             &  2355.5&      22(6)& 2355.7&   0.08&   \\
  0.07&  2821.0(4)&  2377.5&      17( 6)&       &        & \\
       &             &  2149.6&      19(6)& 2149.9&   0.13&   \\
       &  2842.3(5)&  2304.5&      14( 5)&       &        & \\
 0.05&  2854.8(3)&  2854.8&      39(8)& 2854.2&   0.16&   \\
       &             &  2186.3&      12(3)& 2186.7&   0.36&  2\\
  0.06&  2861.6(3)&  2861.6&      42( 8)&       &        & \\
  0.05&  2870.8(6)&  2333.0&      14(5)& 2333.1&   0.31&  2\\
       &  2875.3(4)&  2875.3&      20(6)& 2873.6&   0.05&   \\
       &  2881.4(12)&  2845.9&      16(5)& 2846.3&   0.23&   \\
       &             &  2343.6&      16(5)& 2345.2&   0.07&   \\
       &  2888.6(5)&  2350.8&      16(5)& 2351.8&   0.16&   \\
       &             &  2217.2&      8(3)& 2218.4&   0.10&  2\\
  0.08&  2897.7(4)&  2897.7&      16(5)& 2898.0&   0.16&   \\
       &             &  2434.4&      14(5)& 2435.0&   0.08&  3\\
  0.04&  2907.9(4)&  2370.1&      18( 5)&       &        & \\
  0.10&     2916.2(2)&  2916.2&      86(12)&       &        & \\
  0.18&  2920.0(12)&  2920.0&      40(9)& 2922.3&   0.05&   \\
       &             &  2884.5&      46(8)& 2885.0&   0.05&   \\
       &             &  2476.5&      25( 7)&       &        & \\
       &             &  2456.7&      16( 5)&       &        & \\
       &             &  2248.6&      09( 3)&       &        & \\
       &             &  2190.8&      21(5)& 2190.7&   0.21&  3\\
       &  2933.6(4)&  2933.6&      24( 6)&       &        & \\
  0.08&  2936.8(3)&  2901.3&      34( 7)&       &        & \\
  0.16&  2951.7(4)&  2916.2&      29(7)& 2914.1&   0.08&  2\\
       &  2965.1(5)&  2965.1&      17( 6)&       &        & \\
  0.07&  2972.1(10)&  2972.1&      22( 6)&       &        & \\
       &             &  2434.3&      25( 6)&       &        & \\
       &             &  2939.1&      26(7)& 2940.0&   0.05&   \\
  0.24&  2990.8(6)&  2990.8&      55(9)& 2990.0&   0.12&   \\
       &             &  2955.3&      26( 6)&       &        & \\
       &             &  2527.5&      50(9)& 2527.2&   0.24&   \\
       &             &  2319.4&      10(4)& 2317.9&   0.09&   \\
       &             &  2261.6&      16(5)&       &        & \\
\hline \end{tabular} \end{center}   \newpage%\vspace*{9.8mm}
\begin{center}
\begin{tabular}{|l|r|r|r|r|r|c|}  \hline
$I_1$,\%&  $E_i$, keV &$E_2$, keV &
$i_{\gamma\gamma}$ & $E_s$, keV & $I_s$, \%& M\\\hline   
  0.25&  3002.2(4)&  3002.2&      26( 7)&       &        & \\
       &             &  2966.7&      35( 7)&       &        & \\
       &             &  2538.9&      97(11)&       &        & \\
       &             &  2464.4&      12( 4)&       &        & \\
       &             &  2330.8&      22( 4)&       &        & \\
       &             &  2273.0&      35(6)& 2273.4&   0.12&   \\
  0.25&  3013.0(5)&  2549.7&      18(5)& 2550.8&   0.39&  3\\
       &             &  2475.2&      13(4)& 2477.0&   0.10&   \\
       &             &  2569.5&      33(8)& 2568.2&   0.20&   \\
       &             &  2981.1&      34(7)& 2980.4&   0.19&  2\\
  0.58&  3021.1(4)&  3021.1&      49(7)& 3022.3&   0.09&   \\
       &             &  2985.6&      85(11)& 2986.4&   0.18&  2\\
       &             &  2577.6&      61(11)& 2577.9&   0.13&  2\\
       &             &  2557.8&      59(9)& 2557.4&   0.22&  2\\
       &             &  2483.3&      49(8)& 2483.8&   0.06&   \\
       &             &  2349.7&      13(4)& 2348.2&   0.06&   \\
       &             &  1700.0&      49(11)&       &        & \\
  0.27&  3070.1(3)&  2606.8&      28( 6)&       &        & \\
       &             &  2400.4&      20( 4)&       &        & \\
       &             &  3073.5&      15( 4)&       &        & \\
       &             &  2630.0&      24(8)& 2630.6&   0.06&   \\
       &             &  2610.2&      20( 6)&       &        & \\
       &  3077.9(5)&  3077.9&      27(5)& 3078.0&   0.10&   \\
       &             &  3042.4&      13(5)& 3043.7&   0.12&  2\\
       &             &  2406.5&      10(4)& 2408.3&   0.11&  2\\
       &  3088.3(7)&  3052.8&      12( 4)&       &        & \\
       &             &  2416.9&      18( 4)&       &        & \\
       &  3098.5(3)&  2369.3&      22( 5)&       &        & \\
  0.61&  3106.1(4)&  3106.1&      94(9)& 3106.4&   0.25&  2\\
       &             &  3070.6&      11( 5)&       &        & \\
       &             &  2662.6&    153(16)& 2662.8&   0.11&   \\
       &             &  2642.8&      50(8)&       &        & \\
       &             &  2568.3&      43(7)& 2568.7&   0.20&   \\
       &             &  2434.7&      38(5)& 2435.0&   0.08&  3\\
       &             &  2376.9&      19(4)& 2378.1&   0.07&  2\\
       &             &  2034.2&      27(7)& 2032.1&   0.09&   \\
  0.07&  3136.4(5)&  2673.1&      21( 6)&       &        & \\
       &             &  2598.6&      12( 4)&       &        & \\
       &             &  2408.9&      11(4)& 2408.3&   0.11&  2\\
  0.43&  3142.8(11)&  3142.8&      41(7)& 3142.5&   0.15&   \\
       &             &  3107.3&      62(9)& 3106.4&   0.25&  2\\
       &             &  2617.4&      17(5)& 2616.2&   0.09&  2\\
       &             &  2605.0&      23( 5)&       &        & \\
       &             &  2413.6&      13( 4)&       &        & \\
       &             &  2070.9&    156(14)& 2070.7&   0.31&   \\
       &             &  1821.7&      33(11)&       &        & \\
\hline \end{tabular} \end{center} \newpage%\vspace*{9.8mm}
\begin{center}
\begin{tabular}{|l|r|r|r|r|r|c|}  \hline
$I_1$,\%&  $E_i$, keV &$E_2$, keV &
$i_{\gamma\gamma}$ & $E_s$, keV & $I_s$, \%& M\\\hline   
       &  3149.7(5)&  2624.3&      13( 5)&       &        & \\
  0.48&  3172.9(8)&  2729.4&      26(8)& 2730.6&   0.31&  4\\
       &             &  2709.6&      15( 5)&       &        & \\
       &             &  2443.7&      17( 4)&       &        & \\
       &             &  2502.8&      16(4)& 2500.5&   0.04&   \\
       &             &  2103.1&      47( 8)&       &        & \\
       &             &  2733.4&      29( 8)&       &        & \\
  0.24&  3184.2(6)&  3184.2&      46(8)& 3184.2&   0.15&   \\
       &             &  2658.8&      19(6)& 2659.3&   0.06&   \\
       &             &  2646.4&      12( 4)&       &        & \\
       &             &  2455.0&      25(5)& 2454.2&   0.10&   \\
  0.07&  3190.7(5)&  2747.2&      24( 7)&       &        & \\
  0.11&  3195.0(4)&  3195.0&      17(5)& 3192.7&   0.05&   \\
       &             &  2465.8&      13( 4)&       &        & \\
  0.54&  3208.2(4)&  3208.2&      96(11)& 3207.8&   0.14&   \\
       &             &  2764.7&      55( 9)&       &        & \\
       &             &  3173.1&      37(7)& 3174.1&   0.09&   \\
       &             &  2745.3&      40(12)&       &        & \\
       &             &  2479.4&      18(5)& 2480.1&   0.21&   \\
       &             &  1887.5&   100(16)& 1888.4&   0.15&   \\
  0.17&  3232.7(4)&  2694.9&      40( 6)&       &        & \\
       &             &  1911.6&      45(13)& 1910.5&   0.14&   \\
       &             &  2771.0&      15(5)& 2770.5&   0.10&  2\\
  0.13&  3237.4(7)&  3237.4&      25(6)& 3238.4&   0.10&   \\
       &             &  3201.9&      19( 6)&       &        & \\
  0.13&  3257.8(4)&  2814.3&      31(7)& 2814.2&   0.08&   \\
  0.12&  3269.0(5)&  2731.2&      15( 5)&       &        & \\
       &  3271.2(5)&  2827.7&      21( 7)& 2827.7&   0.20&  2\\
  0.15&  3277.6(13)&  3277.6&      83(10)& 3278.4&   0.29&   \\
       &             &  3242.1&      22(6)& 3242.0&   0.07&   \\
       &             &  2739.8&      17(5)& 2740.8&   0.13&   \\
       &             &  2548.4&      27(5)& 2547.2&   0.06&   \\
       &             &  1956.5&      83(15)&       &        & \\
  0.29&  3292.6(10)&  2767.2&      18(6)& 2767.3&   0.05&  2\\
       &             &  1971.5&      74(15)& 1971.6&   0.11&   \\
       &  3295.9(5)&  2852.4&      24( 7)&       &        & \\
  0.09&  3305.4(5)&  2861.9&      20( 7)&       &        & \\
       &             &  2634.0&      11(4)& 2633.5&   0.10&   \\
  0.15&  3345.8(14)&  3345.8&      23(6)& 3345.3&   0.05&   \\
       &             &  3310.3&      29(8)& 3311.9&   0.13&   \\
       &             &  2902.3&      42(9)& 2902.4&   0.12&   \\
       &             &  2674.4&      13( 4)&       &        & \\
  0.34&  3350.4(12)&  3350.4&   101(11)& 3350.3&   0.16&   \\
       &             &  3314.9&      22( 7)&       &        & \\
       &             &  2906.9&      34(9)& 2906.1&   0.05&  2\\
\hline \end{tabular} \end{center} \newpage%\vspace*{9.8mm}
\begin{center}
\begin{tabular}{|l|r|r|r|r|r|c|}  \hline
$I_1$,\%&  $E_i$, keV &$E_2$, keV &
$i_{\gamma\gamma}$ & $E_s$, keV & $I_s$, \%& M\\\hline
       &             &  2812.6&      29( 6)&       &        & \\
       &             &  2679.0&      10(4)& 2678.4&   0.27&   \\
  0.04&  3387.4(4)&  2849.6&      20( 6)&       &        & \\
     6&  3400.7(7)&  2957.2&      32(9)& 2956.1&   0.50&  2\\
       &             &  2862.9&      18( 5)&       &        & \\
       &             &  2729.3&      16(4)& 2730.6&   0.31&  4\\
       &             &  2671.5&      22(6)& 2670.6&   0.19&  2\\
  0.15&  3404.5(5)&  2675.3&      18(6)& 2674.2&   0.35&  2\\
       &             &  2333.4&      16(6)& 2335.8&   0.12&   \\
  0.15&  3425.8(4)&  2754.4&      11( 4)&       &        & \\
       &             &  2696.6&      16(5)& 2696.6&   0.21&   \\
       &             &  2963.5&      47(8)& 2964.6&   0.18&   \\
  0.47&  3430.1(3)&  3430.1&      39(6)& 3432.1&   0.07&   \\
       &             &  3394.6&   249(18)& 3394.5&   0.48&   \\
       &             &  2986.6&      36(11)& 2986.4&   0.18&  2\\
       &             &  2892.3&      24(6)& 2894.0&   0.16&   \\
       &             &  2758.7&      15( 4)&       &        & \\
       &             &  2700.9&      13(5)& 2699.0&   0.26&   \\
  0.06&  3439.1(6)&  3403.6&      20(6)& 3405.6&   0.08&   \\
       &             &  2901.3&      17(6)&       &        & \\
  0.27&  3443.6(6)&  3408.1&      21( 6)&       &        & \\  
       &             & 3000.1&      28( 8)&       &        & \\
       &             &  2980.3&      15(5)& 2980.4&   0.19&  2\\
       &             &  2905.8&      44( 8)&       &        & \\
       &             &  2371.7&      19( 6)&       &        & \\
  0.25&  3462.5(3)&  3019.0&      24( 8)&       &        & \\
       &             &  2791.1&      15(4)& 2792.9&   0.14&   \\
       &             &  2733.3&      20( 5)&       &        & \\
  0.12&  3472.4(4)&  2934.6&      21(6)& 2933.2&   0.20&  2\\
  0.14&  3477.1(12)&  3477.1&      12( 4)&       &        & \\
       &             &  2747.9&      13( 5)&       &        & \\
  0.10&  3491.6(5)&  3048.1&      23( 8)&       &        & \\
       &  3494.6(3)&  2956.8&      23(6)& 2956.1&   0.50&  2\\
  0.24&  3500.7(8)&  3500.7&      86(9)& 3501.6&   0.17&   \\
       &             &  2962.9&      17( 5)&       &        & \\
       &             &  2829.3&      15( 4)&       &        & \\
       &             &  2771.5&      15( 5)&       &        & \\
       &             &  2428.8&      18(5)& 2428.0&   0.08&  2\\
  0.10&  3511.3(6)&  3475.8&      12( 4)&       &        & \\
       &             &  2839.9&      17( 4)&       &        & \\
       &             &  2784.5&      22(5)& 2781.3&   0.09&   \\
\hline \end{tabular} \end{center} \newpage%\vspace*{9.8mm}
\begin{center}
\begin{tabular}{|l|r|r|r|r|r|c|}  \hline
$I_1$,\%&  $E_i$, keV &$E_2$, keV &
$i_{\gamma\gamma}$ & $E_s$, keV & $I_s$, \%& M\\\hline
  0.08&  3532.4(4)&  3532.4&      17( 4)&       &        & \\
       &             &  3496.9&      42(7)& 3496.5&   0.27&   \\
       &             &  3088.9&      87(13)& 3089.1&   0.10&   \\
       &             &  2994.6&      17( 5)&       &        & \\
       &             &  2803.2&      16(5)& 2801.9&   0.12&   \\
       &             &  2211.3&      28(7)& 2211.3&   0.15&   \\
  0.28&  3555.4(5)&  3555.4&      55(7)& 3554.3&   0.25&   \\
       &             &  3519.9&      25( 6)&       &        & \\
       &             &  3092.1&      54(9)& 3092.1&   0.14&   \\
       &             &  2826.2&      14(4)& 2827.7&   0.20&  2\\
  0.12&  3564.2(5)&  3564.2&      37( 7)&       &        & \\
       &             &  3100.9&      22(6)& 3101.5&   0.06&   \\
       &             &  2492.3&      21(6)& 2492.3&   0.17&   \\
       &  3568.3(5)&  2247.2&      20(6)& 2247.2&   0.13&  2\\
  0.12&  3579.2(6)&  3135.7&      31(11)&       &        & \\  
       &             &  3041.4&      18(6)& 3043.7&   0.12&  2\\
       &             &  2907.8&      12(4)& 2906.1&   0.05&  2\\
       &             &  2850.0&      15( 4)&       &        & \\
  0.07&  3591.2(3)&  2862.0&      20(5)& 2862.7&   0.25&   \\
  0.18&  3605.6(4)&  2284.5&      32(9)& 2283.2&   0.13&  2\\
       &             &  2934.5&      29(5)& 2935.3&   0.25&   \\
  0.25&  3634.1(4)&  2313.0&      31(8)& 2313.7&   0.34&   \\
  0.09&  3645.1(5)&  2915.9&      13(4)& 2916.9&   0.24&   \\
  0.24&  3652.6(3)&  2331.5&      41(8)& 2330.3&   0.17&  2\\
  0.12&  3668.3(3)&  2596.4&      24(5)& 2597.5&   0.16&   \\
  0.17&  3692.0(3)&  2370.9&      46(8)& 2370.2&   0.23&  2\\
  0.25&  3707.1(5)&  2386.0&      26(8)& 2383.1&   0.13&   \\
       &  3717.8(5)&  2645.9&      14(5)& 2644.6&   0.17&  2\\
       &  3718.4(5)&  2397.3&      23(8)& 2396.9&   0.15&   \\
  0.20&  3743.5(4)&  2422.4&      28(8)& 2422.5&   0.15&   \\
  0.05&  3801.2(3)&  2729.3&      26(5)& 2730.6&   0.31&  4\\
  0.19&  3876.5(5)&  2555.4&      29(8)& 2554.7&   0.15&  2\\
  0.27&  3891.2(5)&  2570.1&      27( 9)&       &        & \\
       &  3929.4(5)&  2608.3&      25(9)& 2607.0&   0.19&  2\\
\hline \end{tabular} \end{center}

\newpage
\begin{figure}%[htbp]
\vspace{2cm}
\leavevmode
%\hspace{-.8cm}
\epsfxsize=14cm

\epsfbox{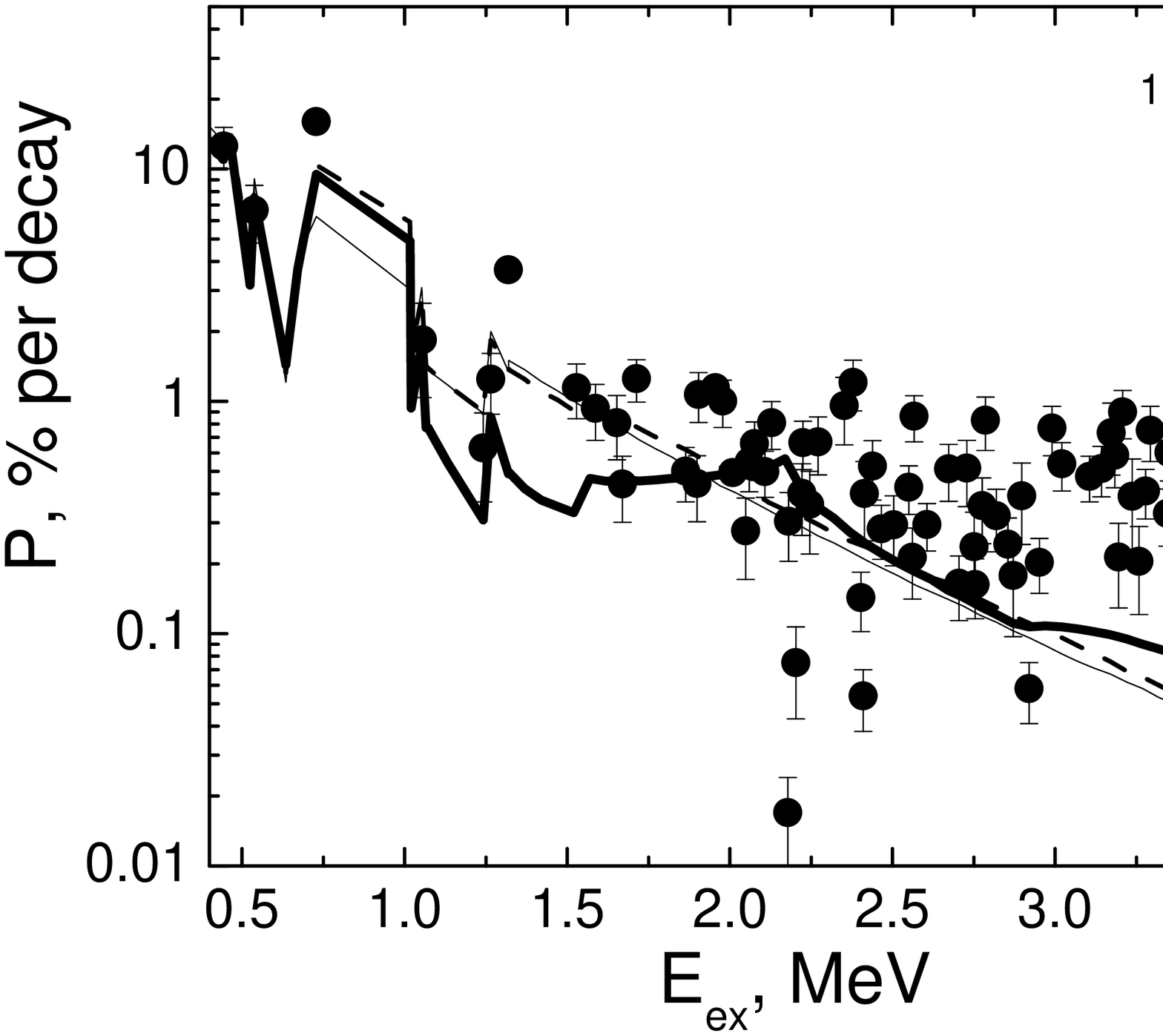}
\vspace{-4.cm}

{\bf {Fig. 1.} \it ~The total population of two-step cascades intermediate
 levels (points  with bars), thin curve represents calculation within models 
[12,14]. Dashed  curve shows results of calculation using data [3]. Thick
curve shows results of calculation using level density [3], and corresponding 
strength functions of secondary transitions are multiplied by function $h$ set 
by equations (1) and (2).}\\\vspace{10.cm}
\newpage
\end{figure}

\begin{figure}[htbp]
\vspace{2cm}
\leavevmode
%\hspace{-.8cm}
\epsfxsize=14cm

\epsfbox{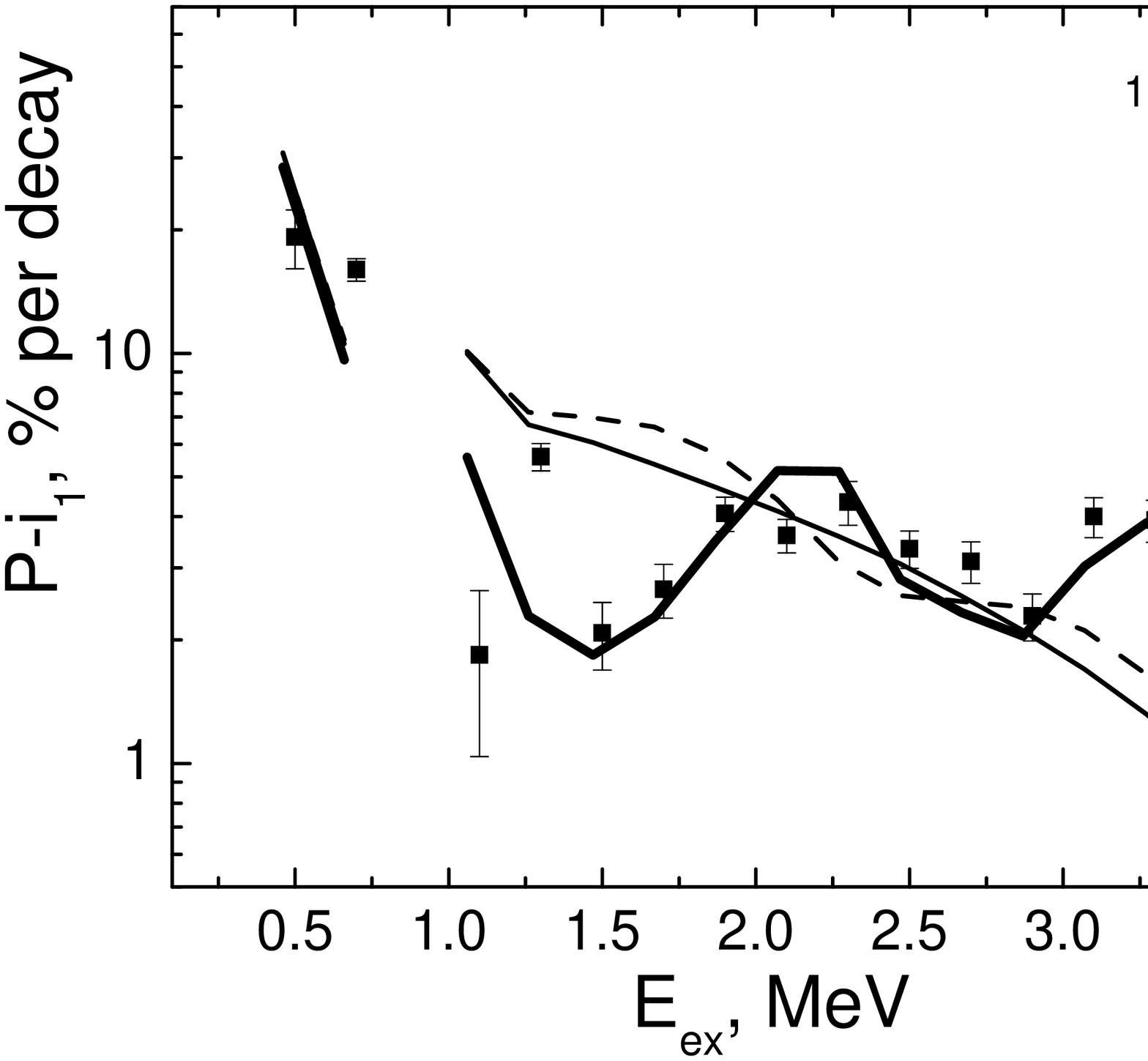}
\vspace{-4.cm}

{\bf {Fig. 2.}\it ~The same, as in Fig.~1, for the cascade population of levels 
in the  200 keV energy bins.}\\
\vspace{10.cm}
 \end{figure}
\newpage
\begin{figure}%[htbp]
\vspace{2cm}
\leavevmode
%\hspace{-.8cm}
\epsfxsize=14cm

\epsfbox{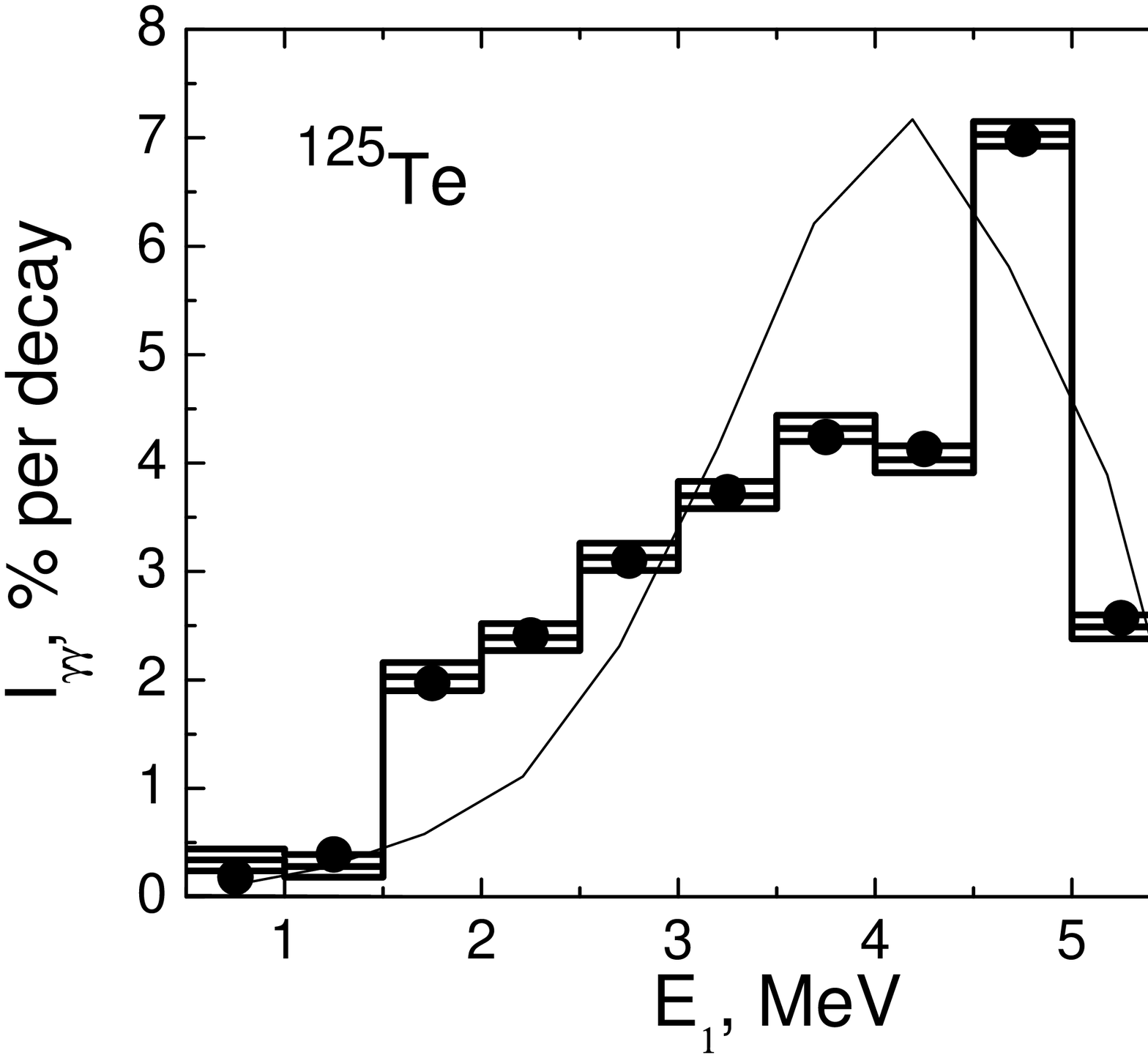}
\vspace{-4.cm}

{\bf {Fig. 3.}\it ~Histogram is the intensity of the two-step cascade in 
function of the energy of their
 primary transitions with statistical errors only. Line is the calculation in 
frame of models [12,14]. Points are the typical fit by the most probable $\rho$ 
and $k$}.\\\\\vspace{10.cm}
\end{figure}
\newpage 
\begin{figure}%[htbp]
\vspace{4cm}
\leavevmode
%\hspace{-.8cm}
\epsfxsize=14cm

\epsfbox{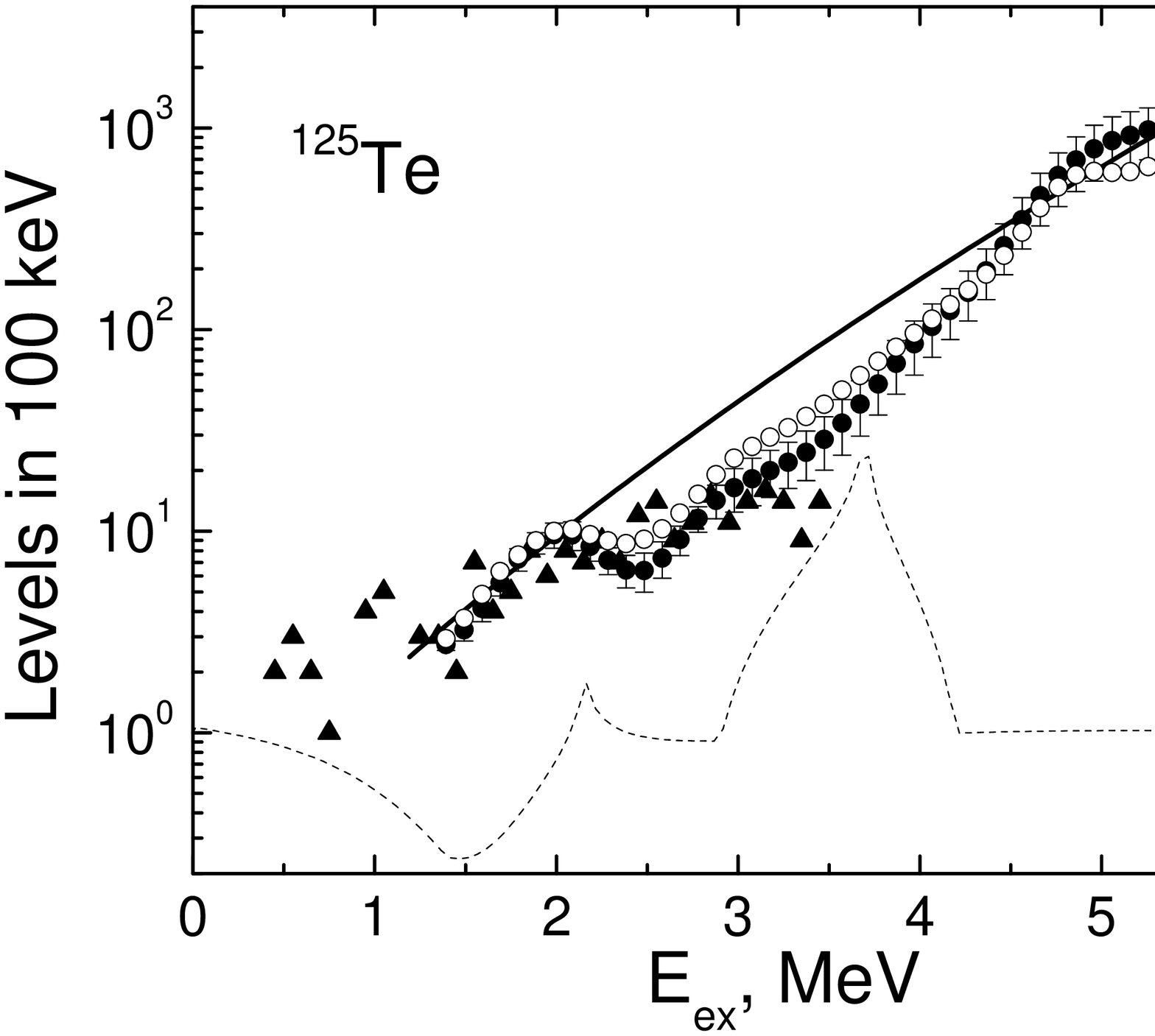}
\vspace{-4cm}

{\bf{Fig. 4.} \it  ~The number of intermediate levels of two-step cascades in 
the case of 
different functional dependence of strength functions for  primary and secondary 
cascade transitions. Dashed line shows values of function $h$ for excitation 
energy
 $B_n-E_1$. Solid line represents predictions according to model [12].
 Triangles are the number intermediate levels of obtained two-step cascades. 
Open points - data from analyses [3].}\vspace{10.cm}
\end{figure}
\newpage
\begin{figure}%[htbp]
\vspace{2cm}
\leavevmode
%\hspace{-.8cm}
\epsfxsize=14cm

\epsfbox{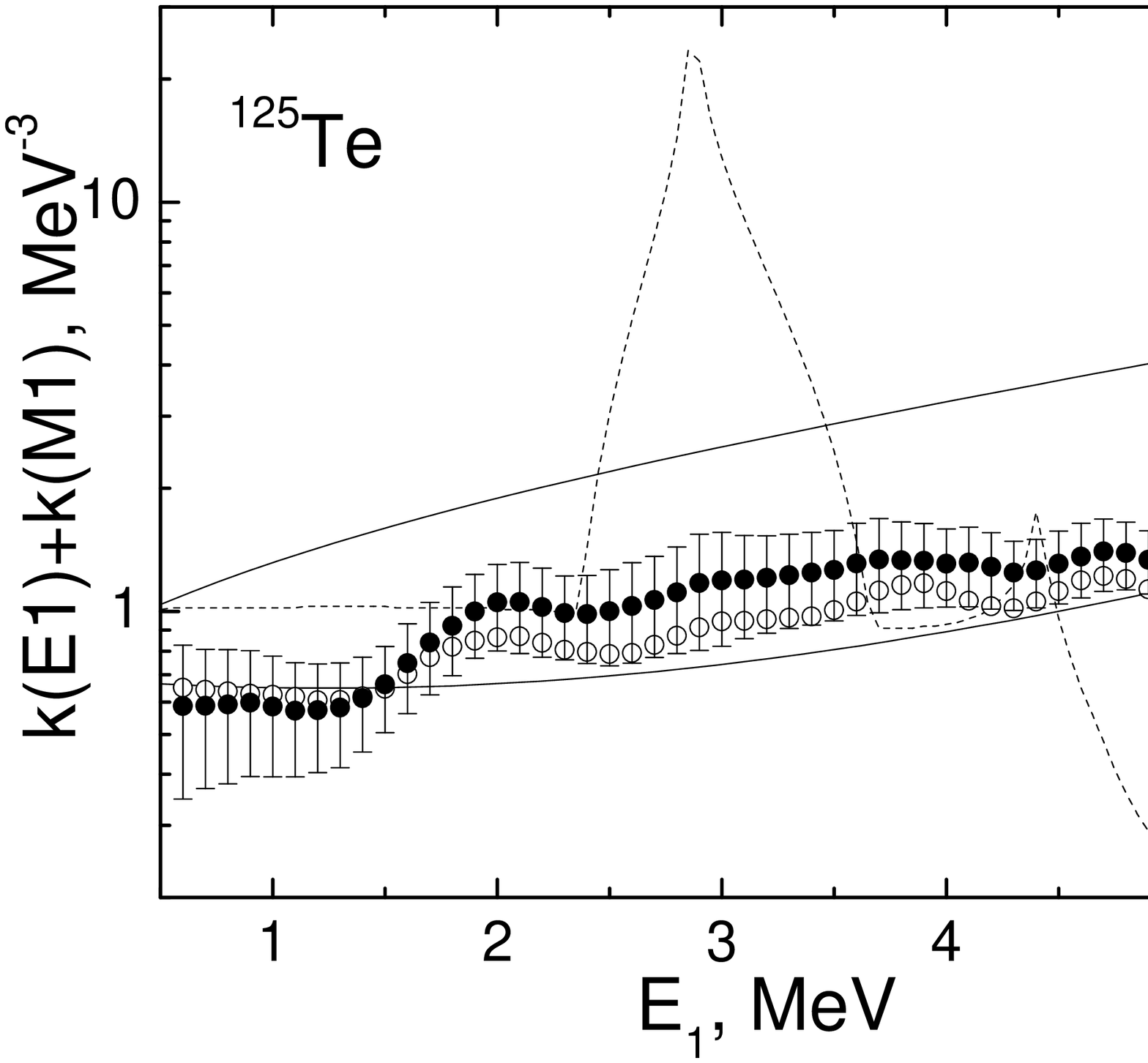}
\vspace{-4.cm}

{\bf{Fig. 5.} \it ~The sums of radiative strength functions of the cascade 
primary dipole
transitions providing reproduction of cascade intensities with the considered 
difference of their values with strength functions of secondary transitions 
(multiplied by $10^9$). Dashed line shows values of function $h$ for excitation 
energy $B_n-E_1$. Solid lines are the models [13,14] predictions with 
$k(M1)=const$. Open points - data from analysus [3].}\\\vspace{10.cm}
\end{figure}
 
\end{document}